\documentclass[twocolumn,showpacs,preprintnumbers,amsmath,amssymb]{revtex4}

\usepackage{graphicx}
\usepackage{dcolumn}
\usepackage{bm}

\begin{document}

\preprint{}

\title{Parameterization for $q_T$-spectrum of inclusive $Z$-boson hadroproduction}

\author{M.V.~Bondarenco}
 \email{bon@kipt.kharkov.ua}
\author{S.T.~Lukyanenko}
\author{P.V.~Sorokin}
 \affiliation{NSC Kharkov Institute of Physics and Technology, 1 Academic St.,
61108 Kharkov, Ukraine}

\date{\today}

\begin{abstract}
We propose a simple parameterization for $q_T$-spectrum of inclusive
$Z$-boson production, and test it against world collider data at
different energies. The fit gives good agreement with the data, and
indicates the existence of two distinguishable breaks in the
$q_T$-spectrum. Energy dependences of the fitted parameters are
discussed.


\end{abstract}





\maketitle


Momentum distributions of heavy bosons ($W^{\pm}$, $Z$, or highly
virtual photons) inclusively produced in hadron-hadron collisions \cite{DY-reviews,Ellis-Stirling-Webber} are widely used for studies of the underlying parton interaction subprocesses, as well as parton distribution functions (pdfs) 
\cite{Dittmar,pdfs-DY}. At high collision energy $\sqrt s$, the
dependences on the boson transverse momentum $q_T$ (with respect to
the hadron collision axis) and on the boson rapidity $Y$ correlate
only weakly, so it makes good sense first to study
single-differential normalized cross-sections
$\frac1{\sigma}\frac{d\sigma}{dq_T}$ and
$\frac1{\sigma}\frac{d\sigma}{dY}$ independently. In this Letter, we
will discuss some trends emerging in the behavior of the
$q_T$-spectrum at highest accessible $\sqrt s$.

As has long been known, $q_T$-spectra of heavy boson production
contain several distinct kinematic regions. In the hard region
$q_T\gtrsim M_V$, with $M_V$ the boson (or dilepton) mass,
perturbative calculations are expected to be reliable. In the
Sudakov region, where $q_T\ll M_V$, partonic differential
cross-sections are enhanced, favoring resummation to all orders of
perturbation theory \cite{DDT,resum}. That is equivalent to account
of parton cascading effects. Finally, there is a soft and genuinely
non-perturbative region $q_T\sim1$ GeV, where hadronic degrees of
freedom in cascading become essential.

At present, the theoretical framework for treatment of the boson
$q_T$-spectrum simultaneously in all regions reached high degree of
sophistication \cite{CSS,Ellis-Veseli,ResBos,generators}, involving
a host of adjustable parameters entering the pdfs, the factorization
and renormalization schemes, and infrared regulators. On the other
hand, the experimental shape of the boson $q_T$-spectrum does not
exhibit many features, so, it can tightly constrain only a few
parameters or their combinations.

Our observation made when working with high-statistics data
(specified below) is that in region $q_T\ll M_V$, the spectrum
basically follows a power-law pattern, whereas at asymptotic $q_T\gg
M_V$, it changes to a power law with a greater spectral index. Even
though there are no firm theoretical reasons for a strictly
power-like behavior anywhere\footnote{Strictly speaking, at $q_T\ll
M_V$ one expects an exponentiated double-logarithmic asymptotics of
the $q_T$-spectrum \cite{resum}, whereas at $q_T\gg M_V$, the
$q_T$-spectrum is proportional to the underlying pdfs, with their
$x$'s proportional to $q_T$. The pdfs at small $x$ (if $\sqrt s\ggg
q_T$) may also have exponentiated double-logarithmic asymptotics
\cite{pdf-DLLA,Dittmar}, though can as well admit power-law (Regge)
parameterizations \cite{Dittmar}.}, it may serve as a fair
approximation in an available limited range of $q_T$. In total, two
power laws contain 2 spectral indices and 2 scales (or
normalizations), hence the minimal number of parameters required to
describe the complete boson production $q_T$-spectrum equals 4.

The transition between the intervals of power-law behavior is
smooth, yet it may be smeared out by poor statistics data. But
nowadays, high-statistics data are delivered by the Tevatron and the
LHC, which may provide a sharper picture of this transition.

In the present Letter, we propose a simple 4-parameter model for the
differential cross-section, and test it against the best data
presently available at different energies. Those data include the
Tevatron CDF measurement of $p\bar p\to e^+e^-X$ production at
$\sqrt{s}=1.98$ TeV, with 2.1 fb$^{-1}$ of integrated luminocity
\cite{CDF2012}, and the LHC measurement at $\sqrt{s}=7$ TeV in
channels $p p\to \mu^+\mu^-X,e^+e^-X$, with 36 pb$^{-1}$ of
integrated luminocity \cite{CMS-qT,ATLAS-qT}. To be more certain
about the trends, we add results of the S$\bar p p$S UA2 experiment
at $\sqrt{s}=0.63$ TeV \cite{UA2}, which has ample statistics in
channel $p\bar p\to e\nu X$ (at the $W$ resonance), although
restricted to the region of low $q_T$ only. The $q_T$-spectrum shape
for $W$-boson production is expected to be about the same as for
$Z$, inasmuch as $M_W\approx M_Z$, and furthermore, since in the
covered region $q_T\ll M_W$, the $q_T$-spectrum should have little
sensitivity to the value of $M_W$ (as if $M_W$ was sent to infinity
at $q_T$ fixed). All the above mentioned measurements are
rapidity-inclusive and select events in a broad vicinity of the
electroweak boson mass resonance.


The ansatz we adopt for the $q_T$-spectrum 
reads
\begin{equation}\label{ansatz}
    \frac1{\sigma}\frac{d\sigma}{dq_T}=\frac1{N}\frac{dN}{dq_T}
=\frac{A(a,\kappa,M,\nu)}{(1+q_T^2/a^2)^{1+\kappa}(1+q_T^2/M^2)^{1+\nu}}q_T,
\end{equation}
where $a$, $\kappa$, $M$, $\nu$ are the adjustable parameters.
Factor $q_T$ in (\ref{ansatz}) stems from the correspondence
$\frac{d\sigma}{dq_T}=2q_T\frac{d\sigma}{dq^2_T}$, while
$\frac{d\sigma}{dq^2_T}=\pi\frac{d\sigma}{dq_xdq_y}$ in our axially
symmetric geometry must be an entire function of
$q^2_T=q^2_x+q^2_y$. Parameter $M$ is not to be confused with the
electroweak boson mass, although we expect it to be of the same
order. Numerator $A(a,\kappa,M,\nu)$ in (\ref{ansatz}) is fixed by
the normalization condition $\int_0^\infty
dq_T\frac1{\sigma}\frac{d\sigma}{dq_T}\equiv1$, giving
\begin{subequations}
\begin{eqnarray}\label{A}
    \frac2A&=&\frac{a^2}{1+\kappa+\nu}F\left(1+\nu,1;2+\kappa+\nu;1-\frac{a^2}{M^2}\right)\quad\label{Aa}\\
&\equiv&\frac{a^2}{\kappa}F\left(1+\nu,1;1-\kappa;\frac{a^2}{M^2}\right)\nonumber\\
&\,&+\frac{a^{2(1+\kappa)}M^{2(1+\nu)}}{(M^2-a^2)^{1+\kappa+\nu}}\text{B}(1+\kappa+\nu,-\kappa),\label{Ab}
\end{eqnarray}
\end{subequations}
where $\text{B}$ is the Euler beta function, and $F$ the
hypergeometric function. If $a\ll M$, Eq.~(\ref{Ab}) simplifies to
\begin{equation}\label{A-approx}
    \frac2{Aa^2}=\frac1\kappa\left[1-\left(\frac{a}{M}\right)^{\!2\kappa}\!\frac{\Gamma(1+\kappa+\nu)\Gamma(1-\kappa)}{\Gamma(1+\nu)}\right]+\mathcal{O}\!\left(\!\frac{a^2}{M^2}\!\right)\!,
\end{equation}
where the second term in brackets, albeit formally small if ratio
$a/M\ll1$, at practice can be $\sim0.3$, and thus non-negligible.
Also note that in the formal limit $\kappa\to0$, the second term in
brackets in (\ref{A-approx})  behaves as $1+\mathcal{O}(\kappa)$,
thus canceling the singularity due to overall factor $\kappa^{-1}$.

Algebraic structure (\ref{ansatz}) is intended to arrange a simple
interpolation between limiting cases where certain expectations
about the spectrum behavior can be stated. Those limiting cases are
described below.

At $q_T\ll M$, Eq.~(\ref{ansatz}) simplifies to
\begin{subequations}
\begin{equation}\label{1stbreak}
    \frac1{\sigma}\frac{d\sigma}{dq_T}\simeq\frac{A}{(1+q_T^2/a^2)^{1+\kappa}}q_T.
\end{equation}
Parameter $a$ regularizes the differential cross-section in the
small-$q_T$ region, where there is no reliable \emph{ab initio}
theoretical treatment, anyway. Parameterization of type
(\ref{1stbreak}) had been phenomenologically successful for
description of $q_T$-spectra of light hadrons produced in
high-energy collisions (see, e.g., \cite{Wong-Wilk}), and of
low-mass dilepton pairs at moderate energies \cite{Yoh}, with
parameter $\kappa\sim$ 4--5. In our problem, though, we do not
expect index $\kappa$ to be that high, because $Z$-boson is a
structureless particle (in contrast to secondary hadrons)
, and because at large collider energies, the participating partons
have momentum fractions $x$ far from the valence region where gluon
and antiquark pdfs receive additional suppression \cite{pdfs-DY}.
Specific forms for low-$q_T$ parameterization may vary: in
particular, a thermodynamically-inspired Tsallis
\cite{Tsallis,Hagedorn} form
has been popular:
\begin{equation}\label{Tsallis}
    \frac1{\sigma}\frac{d\sigma}{dq_T}\simeq\frac{A}{(1+q_T/q_0)^{n}}q_T,
\end{equation}
\end{subequations}
but it has a formal deficiency of being non-analytic in
$\bm{q}_T$-plane at $q_x=q_y=0$.\footnote{Owing to this,
(\ref{Tsallis}) gives a somewhat worse fit at small $q_T$ than
(\ref{1stbreak}) (see \cite{Wong-Wilk}).} We refrain from
thermodynamic analogies for our process, assuming it to be governed
by particle cascading in free space.

In the Sudakov region $a\ll q_T\ll M$, both Eq.~(\ref{1stbreak}) and
(\ref{Tsallis}) reduce to a power law
\begin{equation}\label{interm-power-law}
\frac1{\sigma}\frac{d\sigma}{dq_T}\simeq\frac{Aa^{2(1+\kappa)}}{q_T^{1+\kappa}}.
\end{equation}
When the boson production is calculated by pQCD [see subprocesses
(\ref{parton-subprocesses}) below], the leading
logarithmic asymptotics for subprocess (\ref{qg2qV}) would correspond to behavior $\frac{d\sigma_{qg}}{dq_T}\propto\frac{\alpha_s}{q_T}$, while that for subprocess (\ref{qqbar2gV}), to $\frac{d\sigma_{q\bar q}}{dq_T}\propto\frac{\alpha_s}{q_T}\log\frac{M_V}{q_T}$ \cite{logMqT,Ellis-Stirling-Webber}. The latter behavior can mimic power law (\ref{interm-power-law}) with small and positive $\kappa$. On the other hand, resummation of large logarithms to all orders in $\alpha_s$ can manifest itself as small [$\sim\alpha_s(M_V)$] but negative contribution to $\kappa$. We will not contemplate theoretical estimates of $\kappa$ here, leaving it for experiment. In what concerns consequences for parameter $a$, obviously, with the increase of $\sqrt s$, momentum fractions $x$ of participating partons diminish, and the number of parton branchings increases. By analogy with DIS, that should lead to transverse momentum broadening \cite{broadening}. Hence, $a$ is expected to increase with $\sqrt s$. 

\begin{figure}
\includegraphics[width=\columnwidth]{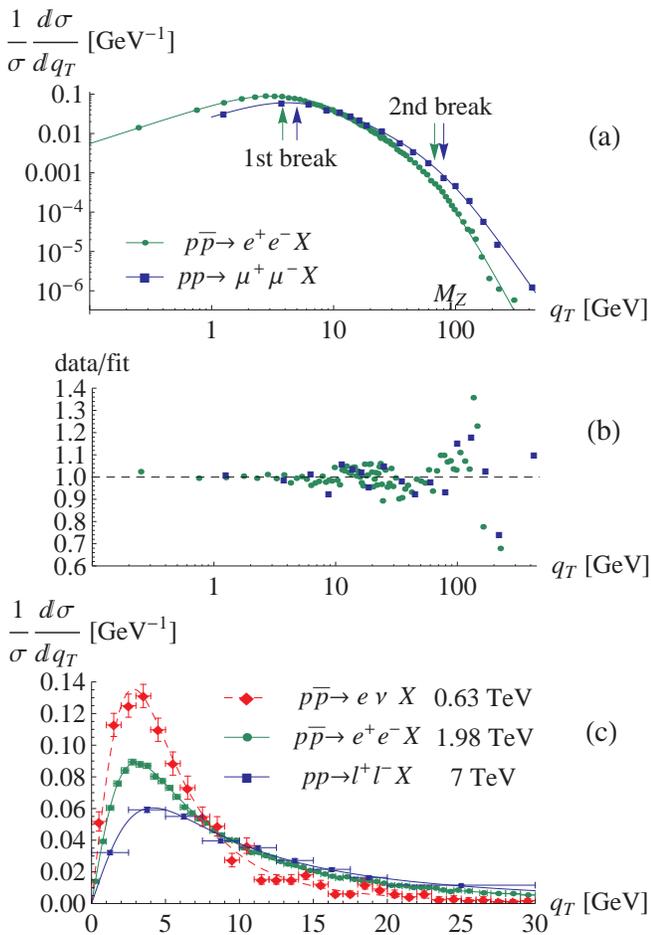}
\caption{\label{fig:LogLog-inset} Comparison of fits to electroweak
boson $q_T$-distributions for different collision energies. (a) The
full experimental $q_T$ range. Green points and the curve, the
Tevatron \cite{CDF2012}; blue points and the curve, the LHC
\cite{CMS-qT}. The arrows show break positions (values of $a$ and
$M$) for each curve. (b) Data to fit ratio for same experiments. (c)
The small-$q_T$ region. Green and blue points and curves, same as in
(a,b). Red points and the dashed curve stand for $W$-boson
production at the S$\bar p p$S \cite{UA2}. }
\end{figure}

On a log-log plot, any power law, in particular, intermediate
asymptotics (\ref{interm-power-law}), and the ultra-low-$q_T$
asymptotics
\begin{equation}\label{simAqT}
\frac1{\sigma}\frac{d\sigma}{dq_T}\simeq A q_T\qquad (\text{at
}q_T\ll a),
\end{equation}
is represented by a straight line. Function (\ref{1stbreak}) then
describes a transition between straight line behaviors of
(\ref{interm-power-law}) and (\ref{simAqT}). It features a break at
$q_T\sim a$, which will be called the 1st break -- see
Fig.~\ref{fig:LogLog-inset}a.


At $q_T\gg M$, Eq.~(\ref{ansatz}) reduces to another power law:
\begin{equation}\label{real-photon-like}
\frac1{\sigma}\frac{d\sigma}{dq_T}\simeq\frac{Aa^{2(1+\kappa)}M^{2(1+\nu)}}{q_T^{3+2(\kappa+\nu)}}.
\end{equation}
Since thereat the boson mass can be neglected, this case must be
close to production of real photons. The transverse energy spectrum
of the latter process is known to fall off as
$\frac1{\sigma^\gamma}\frac{d\sigma^\gamma}{dq_T}\propto q_T^{-5.4}$ for the Tevatron \cite{real-photon-Tevatron}, and $\propto q_T^{-4.5}$ for the LHC \cite{real-photon-LHC}, wherefrom we expect 
\begin{subequations}\label{lambda+nu-LHCTevatr}
\begin{eqnarray}
&\,&\kappa+\nu\approx 1.2\qquad\qquad \text{(Tevatron)},\\
&\,&\kappa+\nu\approx 0.75 \qquad\qquad \text{(LHC).}
\end{eqnarray}
\end{subequations}
Note that physically, the behavior of the massive boson production
differential cross-section at high $q_T$ is sensitive to behavior of
the hadron pdfs, thus, determination of $\kappa+\nu$ at high
$\sqrt{s}$ allows constraining small-$x$ pdf slopes at $Q^2\gtrsim
M^2_V$.

The transition between power-law regions (\ref{interm-power-law})
and (\ref{real-photon-like}) is described by formula
\begin{equation}\label{knee-region}
    \frac1{\sigma}\frac{d\sigma}{dq_T}\simeq\frac{Aa^{2(1+\kappa)}}{q_T^{1+2\kappa}(1+q_T^2/M^2)^{1+\nu}},
\end{equation}
obtained from (\ref{ansatz}) by neglecting the unity in the first
factor in the denominator. Similarly to (\ref{1stbreak}), it
features a break on a log-log plot at $q_T\sim M$, which will be
called the 2nd break (see Fig.~\ref{fig:LogLog-inset}a). The
physical motivation for structure (\ref{knee-region}) is that in the
perturbative region, the boson is expected to be produced chiefly
through $2\to2$ parton reactions
\begin{subequations}\label{parton-subprocesses}
\begin{equation}\label{qg2qV}
\text{quark(antiquark)}+\text{gluon}\to
\text{quark(antiquark)}+\text{boson},
\end{equation}
\begin{equation}\label{qqbar2gV}
\text{quark}+\text{antiquark}\to\text{gluon}+\text{boson},
\end{equation}
\end{subequations}
in which the boson's transverse momentum is balanced by that of a
final parton subsequently transforming into a hadronic jet.
Calculations of those reactions indicate that they indeed form up a
break in the boson's $q_T$-spectrum, although block
$1+q^2_T/M^2_{V}$ enters there in a more complicated fashion than in
(\ref{ansatz}). Such a break has so far not been established
experimentally, so our goal will be to examine whether it manifests
itself is in the new data.

\begin{table}
\caption{\label{tab} The fitted parameter values for experiments
\cite{UA2}, \cite{CDF2012,CMS-qT}. The indicated errors are standard
deviations.}
\begin{ruledtabular}
\begin{tabular}{llllll}
$\!\sqrt{s}$, TeV & $a$, GeV & $\kappa$ & $M$, GeV & $\nu$ & $\kappa+\nu$\\
\hline
0.63 & $5.0\pm0.6$ & $1.1\pm0.1$ & -- & -- & -- \\
1.98 & $3.8\pm0.3$ & $0.29\pm0.04$ & $67\pm5$ & $1.1\pm0.1$ & $1.4\pm0.1$\\
7 & $5.1\pm0.6$ & $0.23\pm0.08$ & $80\pm13$ & $0.54\pm0.09 $ &
$0.8\pm0.1$
\end{tabular}
\end{ruledtabular}
\end{table}


Our approach to handling the data is as follows. Experimental
$q_T$-spectra  of electroweak boson hadroproduction
\cite{CDF2012,CMS-qT,ATLAS-qT,UA2} are published in form of a
cross-section per bin of $q_T$. We ascribe the locations of
experimental points to the bin centres, neglecting the associated
biases. Thereupon, we fit\footnote{To gain sufficient sensitivity of
the fit parameters in the large-$q_T$ region, where the spectrum
falls off by several orders of magnitude, we are actually fitting
the spectrum logarithm.} our model (\ref{ansatz}) to the data points
for experiments \cite{CDF2012,CMS-qT}, and the reduced model
(\ref{1stbreak}) to the low-$q_T$ data of experiment \cite{UA2}.
The overall normalization coefficient $A$ is chosen to obey
Eq.~(\ref{A}), although nothing essentially changes if $A$ is
treated as an independent adjustable parameter.

The fit results are summarized in Figs.~\ref{fig:LogLog-inset} and
Table \ref{tab}. Since the parameterization follows the data
closely, and all the parameters are constrained tightly, our model
may be regarded as phenomenologically reasonable.
The existence of a two-break structure of the $q_T$-spectrum is thus
likely, although there is little room for intermediate asymptotics
(\ref{interm-power-law}). Fig.~\ref{fig:LogLog-inset}b shows the
data to fit ratio as a function of $q_T$. For the LHC experiment,
the deviations seem to be random, whereas for the Tevatron
experiment, there are sign-alternating systematic deviations on the
level of  5--10$\%$, though they are also commensurable with
statistical fluctuations. Naturally, our simple model can not
capture all the physical subtleties; rather, it is surprising that
its systematic deviations are so small.






The obtained parameter values and their differences between the Tevatron and the LHC conditions deserve some physical discussion. First of all, since $a$ and $M$ prove to differ by more than an order of magnitude, the hard and soft scales in our problem are sufficiently well separated. Secondly, it confirms that $M\sim M_Z$, although the actual value of $M$ appears to be somewhat smaller than $M_Z$. 

Next, values of the total spectral index $\kappa+\nu$ in Table~\ref{tab} agree with the differential spectral index for direct photon production [cf. Eqs.~(\ref{lambda+nu-LHCTevatr})]. For the Tevatron, $\kappa+\nu\approx1.4$ is appreciably greater than the corresponding value 0.8 for the LHC. That must be attributed to the fact that at the Tevatron energy, the small-$x$ regime at $q_T>100$ GeV is not reached yet (in contrast to the small-$q_T$ domain), and there is extra suppression of antiquark and gluon distributions in the valence region.  



Index $\kappa$ alone both for the Tevatron and the LHC is small and
positive. Still, it is not quite clear whether it remains $\sqrt
s$-dependent at $\sqrt s\to\infty$. For the S$\bar p p$S experiment,
parameter $\kappa\approx1.1$ is rather large, merely because at this
energy, the partons are characterized by sizable $x\sim M_W/\sqrt s=
0.13$, where antiquark and gluon pdfs experience valence domain
suppressions. But  for the Tevatron and the LHC, the values of
$\kappa$ are commensurable, indicating that at the Tevatron energy,
small-$x$ approximation holds fairly well for moderate $q_T$
already. Hence, the proportion between contributions from reactions
(\ref{qg2qV}) and (\ref{qqbar2gV}) at the Tevatron and the LHC may
be about the same, though not exactly.

Finally, at small  $q_T$, the normalized spectrum is essentially
characterized by parameter $a$ alone (given that $\kappa$ is small
enough). In the past \cite{DY-reviews}, it had been common to
characterize the spectrum width by the mean transverse momentum
$\left\langle q_T\right\rangle=\int_0^\infty dq_T q_T
\frac1{\sigma}\frac{d\sigma}{dq_T}$. However, low-$q_T$
approximation (\ref{1stbreak}) cannot be used for that purpose, as
long as for $\kappa<\frac12$, the spectrum has a too slowly
decreasing `tail', wherewith $\left\langle q_T\right\rangle$
diverges. When evaluated by complete formula (\ref{ansatz}), the
mean transverse momentum equals
\begin{eqnarray}\label{mean-qT}
\left\langle q_T\right\rangle=\frac{a^3A}2\Bigg\{\text{B}\!\left(\frac32,\kappa-\frac12\right)F\!\left(\frac32,1+\nu,\frac32-\kappa,\frac{a^2}{M^2}\right)\,\,\nonumber\\
+\text{B}\!\left(\kappa+\nu+\frac12,\frac12-\kappa\right)\!\left(\frac{M}{a}\right)^{1-2\kappa}\qquad\nonumber\\
\times F\!\left(\kappa+\nu+\frac12,1+\kappa,\kappa+\frac12,\frac{a^2}{M^2}\right)\!\!\Bigg\}\,\,\nonumber\\
\approx a\!\left\{\!\text{B}\!\left(\frac32,\kappa\!-\!\frac12\right)\!+\!\text{B}\!\left(\!\kappa\!+\!\nu\!+\!\frac12,\frac12\!-\!\kappa\!\right)\!\left(\frac{M}{a}\right)^{\!1-2\kappa}\right\}\!.\,\,\,
\end{eqnarray}
If $\kappa$ was greater than $1/2$ (as for the S$\bar pp$S
conditions), the second term in (\ref{mean-qT}) would be
subdominant, owing to $\left({M}/{a}\right)^{1-2\kappa}$ factor.
However, for the Tevatron and LHC conditions, where $\kappa<1/2$, on
the contrary, the second term in (\ref{mean-qT}) dominates and makes
$\left\langle q_T\right\rangle$ a few times greater than $a$.
Therefore, at multi-TeV energies, parameter $a$ is arguably better
suited for characterization of the spectrum width than $\left\langle
q_T\right\rangle$. The maximum of spectrum
$\frac1{\sigma}\frac{d\sigma}{dq_T}$ is achieved at
$q_T\approx\frac{a}{\sqrt{1+2\kappa}}\sim a$, and its height
basically scales as $\sim a^{-1}$, because it equals $\sim Aq_T$
with $A\sim a^{-2}$ [see Eq.~(\ref{A-approx})].

The formidable magnitude of $a$ compared to typical hadron mass
scale is not unusual for semi-inclusive high-energy reactions
\cite{Ellis-Stirling-Webber,CSS,Ellis-Veseli,ResBos,generators}.
It suggests that $a$ may involve a weak dependence on hard scales
$M_V$, $\sqrt s$. In this regard, interesting is the noticeable
difference between the fitted values of $a$ for the Tevatron and the
LHC. That hints that $a$ may slowly increase with $\sqrt s$ starting
from the Tevatron energy (cf. \cite{broadening}). (Therewith,
$\left\langle q_T\right\rangle$ given by Eq.~(\ref{mean-qT}) will
grow, as well.) This is in line with the general trend of the
electroweak boson $q_T$-spectrum broadening continuing since the
S$\bar pp$S (see Fig.~\ref{fig:LogLog-inset}c). Although the value
of $a$ for the S$\bar pp$S is actually greater than for
higher-energy experiments, that may be merely an artifact of much
greater $\kappa$, which, as we argued before, originates from
effects of valence-region suppression at moderate S$\bar pp$S
energy.

We end up with a remark that
%
factor
\begin{equation}\label{alg-factor-a}
    (1+q^2_T/a^2)^{-1-\kappa}
\end{equation}
in (\ref{1stbreak}) with $\kappa\sim0.25$ is reminiscent of the
Cauchy distribution in 2 dimensions \cite{Cauchy-stable}:
\begin{equation}\label{Cauchy2d}
    {(1+q^2_T/a^2)^{-3/2}},
\end{equation}
corresponding to $\kappa=1/2$. The key property of the latter
distribution is its L\'{e}vy-stability under
$\bm{q}_T$-convolutions. Such a stability may be physically relevant
inasmuch as during cascading in the boson production process, the
particles undergo successive transverse momentum redistributions.
Stable distributions exist at $\kappa\neq1/2$ as well, though they
do not assume algebraic form. But rough closeness of $1+\kappa$ to
$3/2$ may explain the phenomenological success of algebraic form
(\ref{alg-factor-a}). Also note that since hadron branching effects
obscure the dependence of $\frac{d\sigma}{dq_T}$ on intrinsic parton
transverse momentum distributions in initial hadrons, 2-scale
factorization of type (\ref{ansatz}) may be more practical than
factoring out the transverse momentum-dependent pdfs and the
resummed hard scattering factor (cf. \cite{DDT,Ellis-Veseli}).


Summarizing, the second break in the $q_T$-spectrum of $Z$-boson
hadroproduction begins to manifest itself at TeV energies, and its
location is close to the $Z$-boson mass. Energy-dependences of
parameters of our parameterization (\ref{ansatz}) are interesting.
In particular, they indicate gradual spectrum broadening and
increase of $a$ with $\sqrt s$, as well as slowdown of $\sqrt
s$-dependences of spectral indices $\kappa$ and $\nu$. The practical
value of extraction of those indices is that $\kappa$ should be
sensitive to parton kinetics in the Regge region, and could be
valuable for resummation studies, while the total spectral index
$\kappa+\nu$ must be closely related to the index of power-law rise
of parton pdfs at small $x$.


\begin{thebibliography}{00}


\bibitem{DY-reviews}

R.~Stroynowski, Phys. Rep. \textbf{71}, 1 (1981);

G.~Altarelli, Phys. Rep. \textbf{81}, 1 (1982).


\bibitem{Ellis-Stirling-Webber}


R.K.~Ellis, W.J.~Stirling, B.R.~Webber. \emph{QCD and Collider
Physics} (Cambridge, Univ. Press, 2003).



\bibitem{Dittmar}
M.~Dittmar \emph{et al.} arXiv:0511119 [hep-ph].

\bibitem{pdfs-DY}

H.-L.~Lai \emph{et al.} Phys. Rev. D \textbf{82}, 074024 (2010);

A.D.~Martin \emph{et al.}
Eur. Phys. J. \textbf{63}, 189 (2009);

R.D.~Ball \emph{et al.}
Nucl. Phys. B \textbf{838}, 136 (2010).







\bibitem{DDT}

Yu.L.~Dokshitzer, D.I.~Dyakonov, and S.I.~Troyan, Phys. Rep.
\textbf{68}, 269 (1980). 

\bibitem{resum}
G.~Parisi and R.~Petronzio, Nucl. Phys. B \textbf{154}, 427 (1979).

\bibitem{CSS}
J.~Collins, D. Soper, and G.~Sterman, Nucl. Phys. B \textbf{250},
199 (1985).

\bibitem{Ellis-Veseli}

R.K.~Ellis and S.~Veseli, Nucl. Phys. B \textbf{511}, 649 (1997).

\bibitem{ResBos}

C.~Balazs and C.-P.~Yuan, Phys. Rev. D \textbf{56}, 5558 (1997).

\bibitem{generators}

A.~Buckley \emph{et al.} Phys. Rep. \textbf{504}, 145 (2011).




\bibitem{pdf-DLLA}

A. De Rujula \emph{et al.} Phys. Rev. D \textbf{10}, 1649 (1974);

R.D.~Ball and S.~Forte, Phys. Lett. B \textbf{335}, 77 (1994).





\bibitem{CDF2012}
T.~Aaltonen \emph{et al.} Phys. Rev. D \textbf{86}, 052010 (2012).

\bibitem{CMS-qT}
S.~Chatrchyan \emph{et al.} Phys. Rev. D \textbf{85}, 032002 (2012).

\bibitem{ATLAS-qT}
G.~Aad \emph{et al.} Phys. Lett. B \textbf{705}, 415 (2011).


\bibitem{UA2}
J.~Alitti \emph{et al.} Zeit. Phys. C \textbf{47}, 523 (1990).


\bibitem{Wong-Wilk}

C.Y.~Wong and G.~Wilk, Phys. Rev. D \textbf{87}, 114007 (2013).


\bibitem{Yoh}

J.K.~Yoh \emph{et al.} Phys. Rev. Lett. \textbf{41}, 684 (1978);

A.S.~Ito \emph{et al.} Phys. Rev. D \textbf{23}, 604 (1981).




\bibitem{Tsallis}

C.~Tsallis, J. Stat. Phys. \textbf{52}, 479 (1988);

S.~Abe, Y.~Okamoto. \emph{Nonextensive Statistical Mechanics and Its
Applications} (Springer-Verlag, 2001).

\bibitem{Hagedorn}

R.~Hagedorn, Riv. Nuov. Cim. \textbf{6}, 1 (1984).

\bibitem{logMqT}

K.~Kajantee and R.~Raitio, Nucl. Phys. B \textbf{139}, 72 (1978).


\bibitem{broadening}

S.~Berge \emph{et al.} Phys. Rev. D \textbf{72}, 033015 (2005).

\bibitem{real-photon-Tevatron}

T.~Aaltonen \emph{et al.} Phys. Rev. D \textbf{80}, 111106 (2009).

\bibitem{real-photon-LHC}
S.~Chatrchyan \emph{et al.} Phys. Rev. D \textbf{84}, 052011 (2011);

G.~Aad \emph{et al.} Phys. Lett. B \textbf{706}, 150 (2011).



\bibitem{Cauchy-stable}

V.V.~Uchaikin, V.M.~Zolotarev. \textit{Chance and Stability. Stable
distributions and their applications} (Utrecht, VSP, 1999).














\end{thebibliography}
\end{document}